\documentstyle[preprint,prb,aps]{revtex} 
\begin{document} 
\draft 
\tighten
\title{Re-entrant resonant tunneling} 
\author{V.V.Kuznetsov, A.K.Savchenko}      
\address
{Department of Physics, Exeter University, Stocker Road, Exeter EX4
4QL, United Kingdom}
\author{M.E.Raikh$^{(1)}$, L.I.Glazman$^{(2)}$} 
\address{
$^{(1)}$ Physics
Department, University of Utah, Salt Lake City, Utah 84112\\   
$^{(2)}$ Theoretical Physics Institute and Department of Physics, 
University of Minnesota, Minneapolis, Minnesota 55455}
\author{D.R.Mace, E.H.Linfield, D.A.Ritchie}  
\address
{Cavendish Laboratory, Madingley Road, Cambridge CB3 0HE,
United Kingdom}

\date{\today}
\maketitle

\begin{abstract} We study the effect of electron--electron interactions on
the resonant--tunneling spectroscopy of the localized states in a barrier.
Using a simple model of  three localized states, we show that, due to the
Coulomb interactions, a {\em single} state  can give rise to {\em two}
resonant peaks  in the conductance as a function of  gate voltage, $G(V_g)$.
We also demonstrate that an additional higher--order resonance with
$V_g$--position in between  these two peaks becomes possibile when
interactions are taken into account. The corresponding resonant--tunneling
process involves two--electron transitions. We have observed both these
effects in GaAs transistor microstructures by studying the time evolution of
three adjacent $G(V_g)$ peaks caused by fluctuating occupation of an
isolated impurity (modulator). The heights of the two stronger peaks exibit
in--phase fluctuations. The phase of fluctuations of the smaller middle peak
is opposite. The two stronger peaks have their origin in the same localized
state, and the third one corresponds to a co--tunneling process.
\end{abstract} 
\pacs{PACS Numbers: 73.20.Dx, 73.40.Gk, 73.40.Sx}


Resonant tunneling  through localized states  in an insulating  barrier (RT)
has been observed in several types of  semiconductor structures. It occurs
when the energy of tunneling electrons  is close to the energy  of a
localized state (LS). If a LS resides near the midpoint of the barrier its
contribution to the conductance is of the order of $e^2/h$. In the first
experiments on diode structures \cite{bend85,nai87,Beasley} the barrier
represented a planar  oxide layer sandwiched between two metallic contacts.
In the later experiments \cite{Geim,geim,AKGeim} a $GaAs$ quantum well with
LS's in it was separated from the contacts by two  $AlGaAs$ barriers. On
small area barriers the authors were able to resolve the tunneling through
individual LS's occuring at resonant bias voltages.

Another group of experiments was carried out  using lateral semiconductor
structures \cite{Fowler,Kopley,McEuen} with a two-dimensional electron gas.
The barrier between the contacts  is created by depleting electrostatically
the region under the gate. The advantage of such a geometry is that the
energy position of a LS with respect to the Fermi level in the contacts can
be varied by changing the gate voltage $V_g$. As $V_g$ is swept, different
LS's in the barrier align with the Fermi level. As a result, narrow peaks
are seen in the $V_g$ dependence of the ohmic conductance $G$. The fact that
the shape of these peaks remains unchanged with decreasing temperature
allows resonant tunneling to be distinguished from the phonon-assisted
transport \cite{Fowler}. The magnitude  of a conductance peak is determined
by the well-known formula \begin{equation} \label{1}
G_{max}=\frac{e^2}{h}\frac{4\Gamma_l\Gamma_r}{(\Gamma_l+\Gamma_r)^2},
\end{equation} where $\Gamma_l$ and $\Gamma_r$ are the tunneling widths of
the LS associated with  electron escape into the left and right contacts
respectively. Since $\Gamma_l,\Gamma_r$ depend exponentially on the location
of the LS within the barrier, the  measurement of the $G(V_g)$ dependence
can be viewed as a resonant--tunneling spectroscopy of localized  states in
the barrier:   the $V_g$ positions of the peaks  reflect the  energies of
LS's,  while their magnitudes characterize the spacial positions of  LS's
with respect to the center of a barrier.

Such a spectroscopy implies that LS's  are well separated in the direction
along the barrier ($y$--direction in Fig.~1). Otherwise, the Coulomb
interaction  between different LS's becomes  important.  Due to this
interaction, the  energy position of a LS, as revealed in  $G(V_g)$
dependence, becomes a function of the occupation of the neighboring LS's. To
illustrate this, consider two LS's with energies  $\varepsilon_1$ and
$\varepsilon_2$ (Fig.~1). Since  $\varepsilon_1 < \varepsilon_2$ the level
$\varepsilon_1$ is the first to ``resonate''  as the Fermi level in the
contacts, $\mu$, moves up with increasing $V_g$. For $\mu > \varepsilon_1$
the level  $\varepsilon_1$ becomes occupied. This causes  an upward  Coulomb
shift,  $U_{12}$, of the second LS. As a result the occupation number of the
second LS will change from $0$ to $1$ not at $\mu=\varepsilon_2$ but at
$\mu=\varepsilon_2+U_{12}$. However, as long as the two--LS's configuration
is considered, one can still attribute each peak in $G(V_g)$ to an
individual LS. In the present paper we demonstrate  that, in general, this
is not the case. Namely we will show that due to the Coulomb interactions a 
{\em single} LS can cause {\em two peaks} in $G(V_g)$ dependence. We call
this effect (resonant tunneling through the same LS at two different
positions of $\mu$)  re--entrant resonant tunneling. We will then present
the experimental evidence for this effect in $GaAs$ transistor
microstructures.

Consider a configuration consisting of three LS's with energies
$\varepsilon_1, \varepsilon_2$ and $\varepsilon_3$ (Fig. 1(a)). Let us trace
the evolution of their occupation numbers with  $\mu$ moving upwards. At
$\mu = \varepsilon_1 = \mu_1$ the first LS becomes occupied. As already
discussed, the second LS changes its occupation at $\mu=\varepsilon_2 +
U_{12}=\mu_2$. Both charged LS's produce the  Coulomb shifts, $U_{23}$ and
$U_{13}$  of the third LS so that with further increasing $\mu$ the third LS
gets charged at $\mu= \varepsilon_3 + U_{13} +U_{23}=\mu_3$. If, as it is
shown in Fig.~1(a), only the first LS is close to the  center of the
barrier, while the other two are displaced towards the left and the right
boundaries, one should expect one strong peak and two weak peaks in 
$G(V_g)$. Note, however, that for above scenario to apply one should verify
that after the charging of the third LS the energies of the first and the
second LS still remain below the Fermi level. Indeed for $\mu>\mu_3$ these
energies acquire positive shifts $U_{13}$ and $U_{23}$ respectively. For the
first LS the corresponding condition $\mu_1 +U_{12}+ U_{13}< \mu_3$ can  be
rewritten as \begin{equation} \label{2} \varepsilon_3 - \varepsilon_1 >
U_{12}-U_{23}. \end{equation} This condition can be violated if the first
and the third LS are close enough in energy. Then the question arises: what
would the evolution of the  occupation numbers be in this case? Note first
that the meaning of inequality (\ref{2}) is that the total energy of two
electrons in the system of three LS's is minimal when they occupy the first
and the second LS. This energy is  equal to $\varepsilon_1 + \varepsilon_2 +
U_{12}$. Consequently, the reversed inequality means that the minimum energy
for  the two electrons corresponds to the occupation of the second and the
third LS and is equal to $E_2=\varepsilon_2 + \varepsilon_3 + U_{23}$. If
this is the case then  the process of the charging of the system will go as
follows. At  $\mu =\mu_1$ the first LS gets charged. At $\mu=\varepsilon_2 +
\varepsilon_3 +U_{23}-\varepsilon_1=\mu^\ast$ we have $\varepsilon_1 +
\mu^\ast = E_2$. At this position of the Fermi level an electron from one of
the contacts and an electron from the first LS occupy the second and the
third LS, leaving the first LS empty. (Note that this position of the Fermi
level is lower than that required to charge just the second LS, since
$\mu^\ast < \mu_2$ if the condition (\ref{2}) is violated.) Finally, the
first LS {\em will be charged again} at $\mu = \varepsilon_1 + U_{12} +
U_{13}> \mu^\ast$.

From the evolution of the occupation numbers described above one should
anticipate two strong resonant--tunneling peaks in $G(V_g)$: at
$\mu=\varepsilon_1$ and at  $\mu = \varepsilon_1 + U_{12} + U_{23}$
corresponding to the resonance of the first LS.  Note however, one more
resonant--tunneling process  is possible at an intermediate value of $\mu$,
namely at $\mu=\mu^\ast$. The ``resonating'' states are the states of two
electrons. They are:

(i) initial: an electron at the Fermi level in the left contact and 
electron on the first LS,

(ii) intermediate: one electron on the second and another on the third LS,

(iii) final: an electron at the Fermi level in the right contact and
electron on the first LS.

\noindent All three energies are equal at  $\mu=\mu^\ast$. It can be shown
that for such a process the magnitude of the peak in the conductance is
still given by Eq.~(\ref{1}), but the values of $\Gamma_l$ and $\Gamma_r$
are now determined by {\em  two--electron} transitions in the following way.

One can write the general expressions for both widths: $\Gamma_l =
\pi|V_l|^2g$, $\Gamma_r = \pi|V_r|^2g$, where $g$ is the density of states
in the  contacts and $V_l$, $V_r$ are the corresponding matrix elements. In
the case of the resonant tunneling through the  first LS one has $V_l =
V_0exp(-\frac{R_{1l}}{a})$, $V_r = V_0exp(-\frac{R_{1r}}{a})$, where $a$ is
the localization radius, $R_{1l}$, $R_{1r}$ are the distances from the first
LS to the left and to  the right contacts respectively, and $V_0$ is the
prefactor. Then the  formula (\ref{1}) for $G_{max}$ takes the form
\begin{equation} \label{3} G_{max} = \frac{e^2}{h} \cosh^{-2}\left(\frac
{R_{1l}-R_{1r}}{a}\right). \end{equation}

For the two--electron transition between the states (i) and (ii) the matrix
element $\tilde V_l$ can be presented as \begin{equation} \label{4} \tilde
V_l = V_0\exp\left(-\frac{R_{l2}}{a}\right)
\left[\frac{V_0\exp(-\frac{R_{13}}{a})} {\mu^\ast - \varepsilon_2 - U_{12}}+
\frac{V_0\exp(-\frac{R_{13}}{a})} {\varepsilon_1 - \varepsilon_3}\right],
\end{equation} where $R_{l2}$ is the distance from the second LS to the left
contact and $R_{13}$ is the distance between the first and the third LS's.
Two terms in the matrix element (\ref{4}) reflect the fact that two steps of
the transition can be carried out in a different order (and thus through 
different virtual states). We emphasize that this matrix element is
non--zero only due to the  interactions. Indeed, it is easy to see that in
the absence of the Coulomb shifts two terms in (\ref{4}) cancel each other
out.

The matrix element $\tilde V_r$ of the transition between the states (ii)
and (iii) has a  form similar to (\ref{4}) \begin{equation} \label{5} \tilde
V_r = V_0\exp\left(-\frac{R_{r3}}{a}\right)
\left[\frac{V_0\exp(-\frac{R_{12}}{a})} {\mu^\ast - \varepsilon_3 - U_{13}}+
\frac{V_0\exp(-\frac{R_{12}}{a})} {\varepsilon_1 - \varepsilon_2}\right],
\end{equation} where $R_{r3}$ is the distance from the third LS to the right
contact and $R_{12}$ is the distance between the first and the second LS's.
Using Eqs.~(\ref{4}), (\ref{5}) one can find the two--electron widths 
$\tilde \Gamma_l = \pi|\tilde V_l|^2g$, $\tilde \Gamma_r = \pi|\tilde
V_r|^2g$. Substituting $\tilde \Gamma_l$, $\tilde \Gamma_r$ into
Eq.~(\ref{1}) we obtain the following expression for the maximal conductance
\begin{equation} \label{6} \tilde G_{max} = \frac{e^2}{h}
\cosh^{-2}\left(\frac {R_{l2}+R_{13}-R_{12}- R_{r3}}{a}+\gamma\right),
\end{equation} where the dimensionless  constant $\gamma$ is defined as
\begin{equation}\label{7}
\gamma=\ln\left|\frac{(U_{12}-U_{23})(\varepsilon_2-\varepsilon_1)
(\varepsilon_2-\varepsilon_1+U_{23}-U_{13})}{(U_{13}-U_{23})(\varepsilon_3-
\varepsilon_1) (\varepsilon_3-\varepsilon_1+U_{23}-U_{12})}\right|.
\end{equation} We see that the condition $\tilde G_{max}\simeq e^2/h$ has
the form $R_{l2}+R_{13}=R_{12}+R_{r3}$, and it can be satisfied even if the
first LS is displaced from the center of the barrier, so that $G_{max}$ is
small.
  
The conduction involving virtual states considered here is similar to the
elastic co--tunneling  process in the Coulomb blockade\cite{cotunn}. In
co--tunneling, electrons traverse a conducting grain using virtual states
when the transition via a real state is forbidden by the charging
energy\cite{cotunn,Matveev}. The main difference is in the number of virtual
states. The quasi--continuum of states available in a metallic grain leads
to the interference suppression of the co--tunneling rate\cite{cotunn},
while such a suppression obviously does not occur in the problem considered
here.

To study the effect of the re--entrant RT experimentally, we used
sub--micron GaAs MESFETs with lateral barriers\cite{Hopping5,Preprint}.
Earlier,  on these structures we demonstrated the presence of elementary RT
channels in $G(V_g)$ and $I(V)$ characteristics and distinguished between
one-- and  two--impurity RT.  This lateral  structure with a smooth barrier,
Fig.~1, is well suited for testing the discussed model.  It  has an enhanced
density of localized states near the contacts which will not give large
peaks in $G(V_g)$ unless two--impurity RT is involved\cite{Preprint},
although a change in the charge of these states with varying $V_g$ can
affect the  states close to the middle of the barrier. By a standard
lock--in technique we have measured the conductance of GaAs MESFETs   with a
donor concentration of $10^{17}$ cm$^{-3}$ . Highly doped source--drain
contacts have metallic--like conduction. The barrier between the source and
drain with  localized states in it is produced by a gate of  length $L$=0.2
$\mu$m in the direction of the current and width $W$=20 $\mu$m in the
perpendicular direction. The height of the barrier is controlled by the gate
voltage which shifts the localized states in the barrier with respect to the
Fermi level in the contacts.

The experimental demonstration of the re--entrant resonant tunneling should
provide an evidence that the two peaks in $G(V_g)$ have their origin in the
same localized state. Firstly, we expect that the heights of these peaks
should be similar. (A possible difference in the heights can be due to some
effect on the values of $\Gamma_l$ and $\Gamma_r$ in (1) when levels 2 and 3
are charged.) Secondly, the external fields should influence both peaks in
the same way, if the peaks originate from the same energy level. If this
field is a random electric field caused by the occupancy fluctuations of an
isolated impurity (modulator), such a modulator can serve as a local probe
to the arrangement of localized states in the barrier. To explore this
possibility we have measured $G(V_g)$ curves repeatedly for a long period of
time  (several hours) and analysed the fluctuations of peaks in $G(V_g)$
with time.

Fig.~2 shows an example of $G(V_g)$ at T=100 mK with the peaks which we
associate with one--impurity RT channels. Among a dozen peaks there is a
group of three neighbouring peaks labelled A, B and C that have two distinct
states at different times. Two--level conductance fluctuations in mesoscopic
structures (Random Telegraph Signal) are commonly connected with  a single
impurity placed close to a  conducting path of small size\cite{Rts}. The
fluctuations of the occupancy of this impurity produce a random electric
field which modulates  the conductance of the path. The surprising feature
of the fluctuations of the three peaks in Fig.~2 is that these fluctuations
are coherent: the decrease of peaks A and C and increase of peak B occur
simultaneously. Fig.~3(a) shows the fluctuations of the peak amplitudes with
time. Comparison of the relative magnitudes of these fluctuations  shows
that for the peaks A and C they are very close: $G_{high}/G_{low}$ is 1.52
for peak A and 1.58 for peak C. The  fluctuations in the amplitude are
accompanied by distinct but less pronounced shifts in the positions of the
peaks, Fig.~3(b). We have also found a similar group of three peaks with
coherent fluctuations on another sample from the same wafer. The only
difference in the second sample is that the fluctuations of the middle peak
B, which has the smallest amplitude, are hardly detectable.

The fact that the observed fluctuations of the peak amplitude are
simultaneous and have the same relative magnitude suggests that the peaks A
and C can be considered as being due to the same LS (state 1 in our model).
Let us   examine first  an alternative  explanation where these peaks are
due to different LSs. We will see that this straightforward view can be
dismissed as highly improbable. Expression (1) for the tunneling peak
amplitude can be reduced to
$G_{max}=(e^2/h)(\Gamma_{min}/\Gamma_{max})=(e^2/h)\exp[-(R_{max}-R_{min})/a]$,
where $\Gamma_{min}(\Gamma_{max})$ is the minimal (maximal) value of
$\Gamma_l$, $\Gamma_r$ and $R_{max}(R_{min}$) is the corresponding distance
from the LS to the contacts, Fig.1,b. Similar values of $G_{max}$ for the
two peaks means that the other LS has to lie on the same line along the
barrier as state 1 (or on the symmetric about the centre line). The equal
effect of the modulator on the two states means that their distances to the
modulator have to be the same. Thus, for a given state 1 there is only one
possible place for the other state to satisfy the experimental facts:  at
the intercept of the sphere with the modulator in its center, with the line
in the 2D plane of the barrier, see Fig.~1(b). The uncertainty in the
position of this state in the plane is of the order of $a^2$, where $a$ is
the localization radius, and in the energy scale it is of the order of
$\Gamma_{max}$.  With an accuracy of a numerical factor, the probability to
meet such a localized state $P_{1-2}$ is  $g_{2D}a^2\Gamma_{max}$, where
$g_{2D}$ is the two--dimensional density of states. In our structures the
value of $g_{2D}$ is about ten times less than the conduction band value
$3\times 10^{10}$ cm$^{-2}$meV$^{-1}$, and $a\approx 100$ $\AA$.  The rate
of the Fermi level shift with gate voltage $\alpha=d\mu/deV_g\approx 0.2$
\cite{Preprint} gives   $\Gamma_{max}\approx 0.2$ meV from the width of the
peaks in the $V_g-$scale. For the probability $P_{1-2}$ we then obtain a
very small value of $6\times 10^{-4}$.

For the re--entrant tunneling  view  we have to estimate the probability to
meet two states in the vicinity of the first LS. These states should lie
within an area of $L_c$x$L_c$ for the Coulomb interaction not to be screened
by the metallic gate, where $L_c$ is the distance between the gate and the
conducting channel ( in our structures $L_c\approx 1000$ $\AA$ and is
comparable to an effective barrier length ). The characteristic energy
interval of these two states should be  now of the order of the Coulomb
energy $U_{ij}\approx e^2/\kappa L_c\approx 1$ meV. This value agrees with
the estimation of $U_{ij}$ from the separation $\Delta V_g$ between peaks A 
and C in Fig.~2, which in the re--entrant tunneling model is
$U_{12}+U_{13}=\alpha e\Delta V_g\approx 1$ meV. The probability $P_{1-23}$
to meet the two impurities is then $(g_{2D}U_{ij}L_c^2)^2\approx 0.1$. The
latter is two orders of magnitude larger than the value of $P_{1-2}$ and it
is large enough to make the effect of re--entrant tunneling observable.

In our explanation the only feature of the modulator that is important is
that it has a local effect on the surrounding localized states.  Its most
probable position is between the gate and the conducting channel as  is
indicated in Fig.~1(b). This conclusion is drawn from the fact that we do
not see incoherence in the fluctuations of the peaks A and C which are
shifted in $V_g$-scale by $\Delta V_g$=6 mV.  It means that this change in
$V_g$ does not affect the average occupancy of the modulator. For an energy
level under the gate such a change in $V_g$ will indeed have a small effect,
as its occupancy is formed under the condition of a strong electric field
between the gate and channel, $V_g\approx -$1.7 V. The fluctuations of the
modulator charge in time can result from a small gate--channel leakage
current, as it was seen in \cite{Cobden}.

We explain the difference in the amplitudes of peaks A and C by the
influence of charged levels 2 and 3 on $\Gamma_{min}$ and $\Gamma_{max}$ in
the expression for the first level resonance
$G_{max}=(e^2/h)(\Gamma_{min}/\Gamma_{max})$. It is interesting to note that
the modulator produces a similar change in the peak amplitudes but without
significant change in the  $V_g$-- position of the peaks (the shift
accompanying the amplitude modulation is ten times less than the separation
between peaks A and C). Within our model, with the second and the third LSs
positioned on the opposite sides of the first LS, this can be understood as
follows. Although states 2 and 3 are closer to the state 1 than the
modulator and thus produce a stronger Coulomb shift of the energy position
of state 1, their effects on $\Gamma_{min}$ and $\Gamma_{max}$ are partially
compensated in $G_{max}$ which is the ratio of the two values.  The
modulator, however, is most probably located assymetrically with respect to
the first LS and it mainly affects one of the two $\Gamma$s, so that the
compensation does not occur.

Peak B in Fig.~2 has a much smaller amplitude than peaks A and C. In
addition, its modulation is in anti--phase with that of the peaks A and C:
the decrease of the peak B corresponds to the increase of the other two.
This implies that its origin is not in the state 1. For the states 2 and 3
we would normally expect to see  {\em two} small peaks, unless these states
are positioned too far from the middle of the barrier to give a measurable
conductance. The presence of only one resonance allows us to suggest that
the peak B in Fig.~2 is due to   {\em co--tunneling} with all states 1, 2
and 3 involved.  Note that for peaks A and C the modulation results from the 
change of the ratio $\Gamma_l /\Gamma_r$ in (1) by the electric field of the
modulator, whereas for co--tunneling (6) it is the ratio $\tilde \Gamma_l
/\tilde \Gamma_r$ that determines the peak amplitude. It is quite plausible
that when one ratio increases the other decreases and vice versa, which 
would lead to the anti--phase modulation of the co--tunneling peak.

In conclusion, we have demonstrated that charging effects can drastically
change  the resonant--tunneling spectroscopy of localized states in the
barrier. These effects cause the doubling of the single--impurity peaks and
also give rise to additional peaks due to many--electron tunneling.

The work at the University of Minnesota was sponsored by NSF Grant
DMR-9423244.

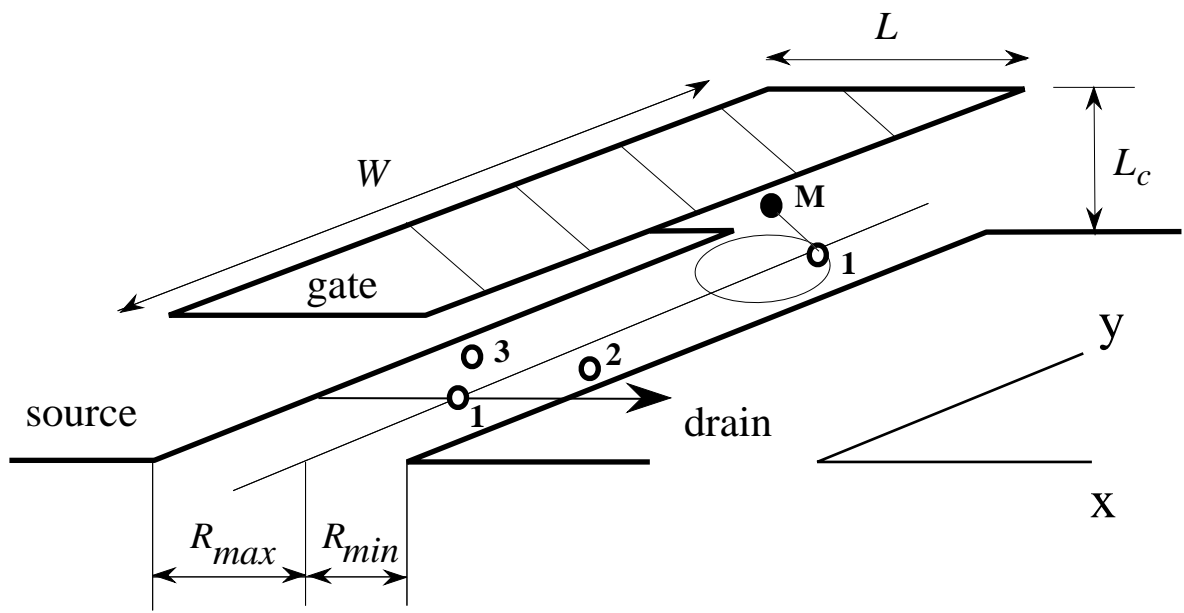
\begin{figure}
\caption{a) Schematic diagram of the barrier with resonant tunneling through
the first localized state (states 2 and 3 are not occupied). b) Spacial
position of the localized states and a modulator in the potential barrier
under the gate. The part with the state 1 and modulator M illustrates the
estimation of the probability for the peak C to be due to an independent
localized state. The part with three localized states (1,2 and 3)
corresponds to the probability to meet three interacting states.}
\label{autonum}
\end{figure}

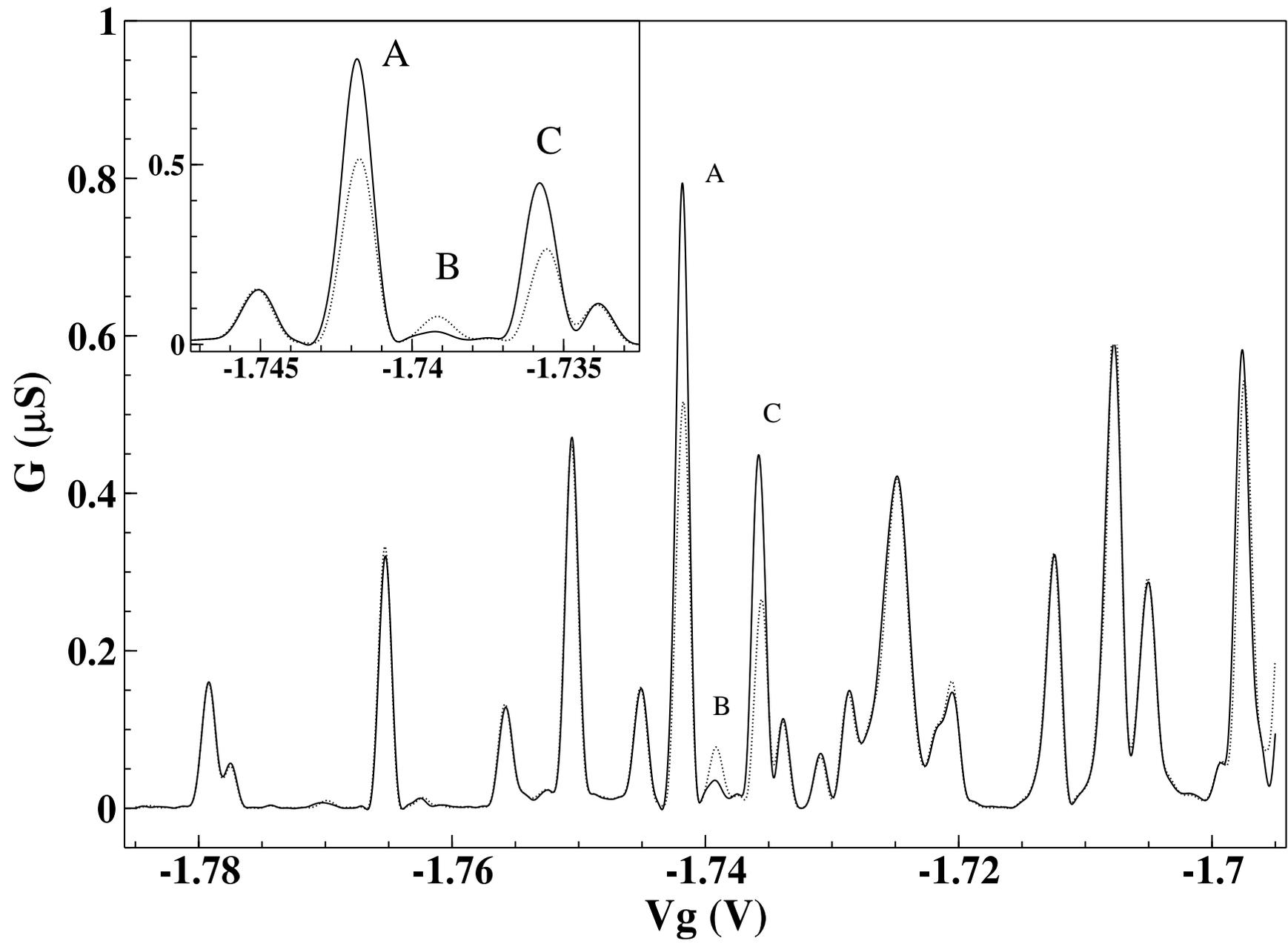
\begin{figure} 
\caption{Two consecutive measurements of the ohmic
conductance as a function of gate voltage. Peaks A,B and C show fluctuations
between distinct two states while the neighbouring peaks are time
independent.} \label{autonum} \end{figure}

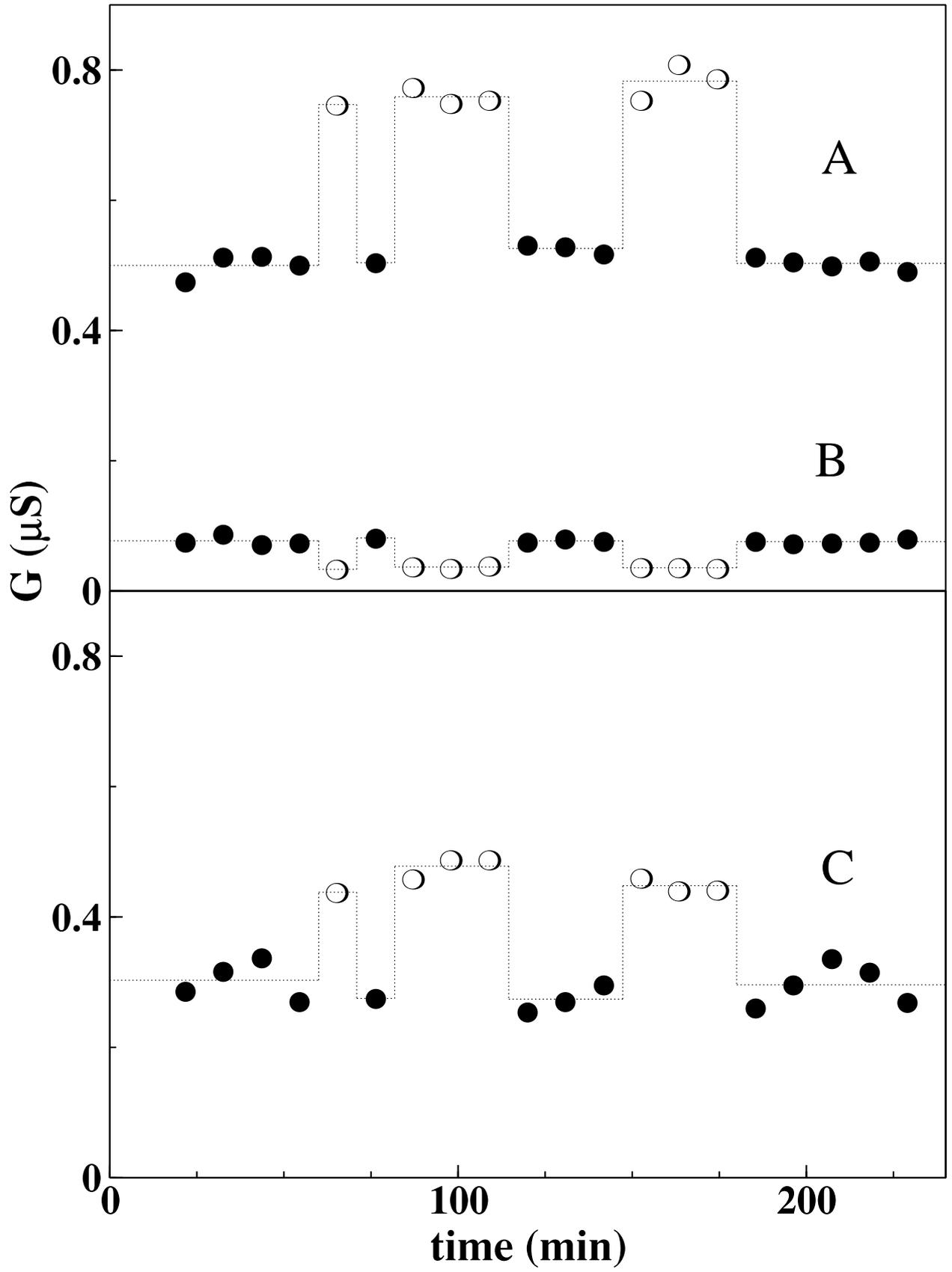
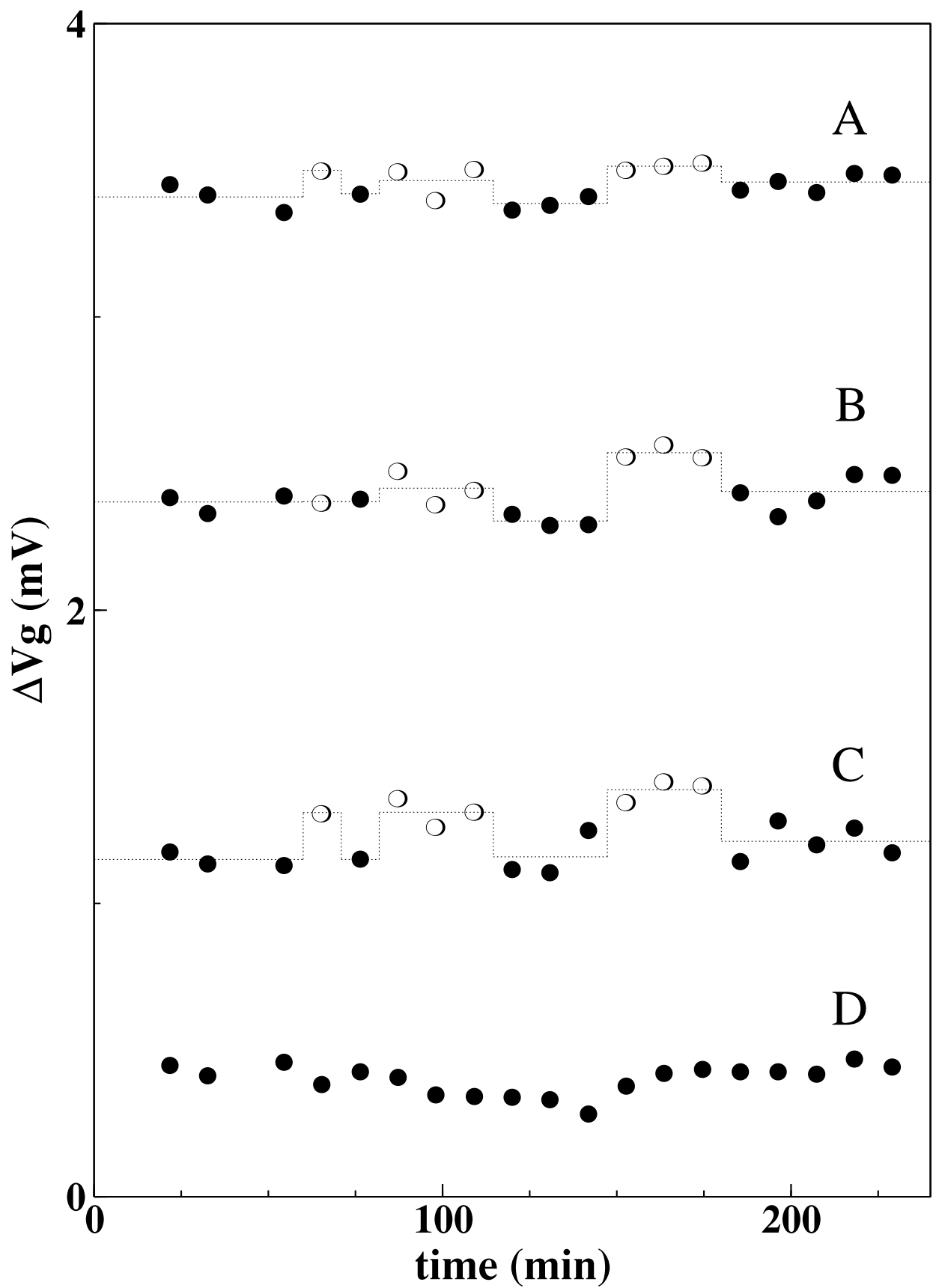
\begin{figure}
\caption{a) Fluctuations of the amplitude of peaks A,B and C with time. b)
Fluctuations of the positions of these peaks in $V_g$-scale accompaning the
amplitude fluctuations. The peak D at $V_g=-1.734$ V is the neighbour of
peak C and it does not show coherent fluctuations.}
\label{autonum}
\end{figure}


\begin{references}

\bibitem{bend85}S. J. Bending and M. R. Beasley, Phys.
Rev. Lett. {\bf 55}, 324 (1985).

\bibitem{nai87} M. Naito and M. R.
Beasley, Phys. Rev. B {\bf 35}, 2548 (1987).

\bibitem{Beasley} Y.Xu, A.Matsuda, M.B.Beasley,
Phys.Rev.B {\bf 42}, 1492 (1990).

\bibitem{Geim} M.W.Dellow, P.H.Beton, C.J.G.M.Langerak, T.J.Foster,
P.C.Main,  L.Eaves, M.Henini, S.P.Beaumont, and C.D.W.Wilkinson,
Phys.Rev.Lett {\bf 68},  1754 (1992)

\bibitem{geim} A.K.Geim, P.C.Main, N.La Scala, Jr., L.Eaves, T.J.Foster,
P.H.Beton,  J.W.Sakai, F.W.Sheard, M.Henini, G.Hill and M.A.Pate,
Phys.Rev.Lett. {\bf 72}, 2061 (1994)

\bibitem{AKGeim} A.K.Geim, T.J.Foster, A.Nogaret, N.Mori, P.J.McDonnell,
N.La Scala, P.C.Main, L.Eaves, Phys. Rev.B 50, N11, 8074-8077 (1994)


\bibitem{Fowler}A. B. Fowler, G. L. Timp, J. J. Wainer, and R.A.Webb,
Phys.Rev.Lett. {\bf 57}, 138 (1986)

\bibitem{Kopley} T.E.Kopley, P.L.McEuen and R.G.Wheeler, Phys.Rev.Lett.
{\bf 61}, 1654 (1988).

\bibitem{McEuen} P.L.McEuen, B.W.Alphenaar, R.G.Wheeler, R.N.Sacks,
Surf.Science 229, 312 (1990)

\bibitem{cotunn} D.V. Averin, Yu.V. Nazarov, Phys. Rev. Lett., {\bf 65},
2446 (1990)

\bibitem{Matveev} L.I. Glazman, K.A. Matveev, Pis'ma Zh.Eksp. Teor. Phys.
{\bf 51}, 425 (1990) [JETP Letters,  {\bf 51}, 484 (1990)]

\bibitem{Hopping5} A.K.Savchenko, A.Woolfe, V.V.Kuznetsov, M.Pepper,
J.E.F.Frost, D.A.Ritchie, M.P.Grimshaw, G.A.C.Jones, in  {\em Hopping and
related phenomena} {\bf 5}, 41, edited by C. J. Adkins, A. R. Long,  and J.
A. McInnes ( World Scientific,  Singapore, 1993)

\bibitem{Preprint} A.K.Savchenko, V.V.Kuznetsov, A.Woolfe, D.R.Mace,
M.Pepper, D.A.Ritchie, and G.A.C.Jones, Phys. Rev.B {\bf52}, N15, R17021
(1995)

\bibitem{Rts} M.J.Kirton and M.J.Uren, Advances in Physics 38, N4, 367-468
(1989)

\bibitem{Cobden} D.H.Cobden, A.Savchenko, M.Pepper, N.K.Patel, D.A.Ritchie,
J.E.F.Frost, and G.A.Jones, Phys.Rev.Lett. {\bf 69}, 502 (1992)

\end{references}
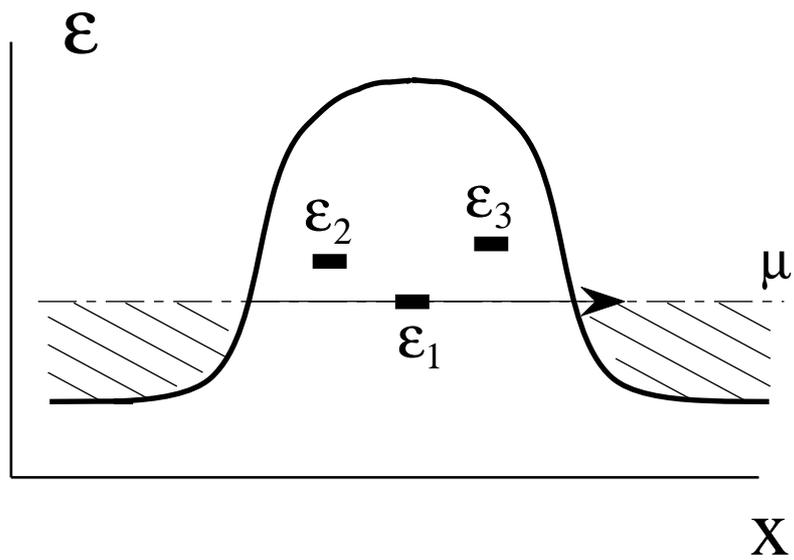
\end{document}